\documentclass[aps,prl,twocolumn,superscriptaddress,nofootinbib]{revtex4-1}

\usepackage{amsmath, amssymb}
\usepackage{slashed}

\newcommand{\Sf}{S_F}
\newcommand{\SfI}{S_F^{-1}}
\newcommand{\dL}{\displaystyle\int\frac{d^d\ell}{(2\pi)^d}}

\begin{document}

\title{The Gauge-Invariant Mass Function}
\author{Kang-Sin Choi}
\email{kangsin@ewha.ac.kr}
\affiliation{Scranton Honors Program, Ewha Womans University, Seoul 03760, Korea}
\affiliation{Institute of Mathematical Sciences, Ewha Womans University, Seoul 03760, Korea}

\begin{abstract}
In gauge theories, the mass of a field has been regarded as a purely on-shell concept: the pole mass is gauge-invariant, but the off-shell propagator has had no gauge-invariant definition of mass. We show that renormalization defines a gauge-invariant mass function at every virtuality, together with a gauge-invariant vertex. The virtual particle becomes as well defined as the on-shell one: the distinction is not dynamical but purely kinematic.
\end{abstract}

\maketitle

Mass is the most basic property of a
particle, yet its meaning in quantum field theory has not yet been
fully clarified. In classical mechanics, mass is a fixed
parameter in the Lagrangian that we can measure without
disturbing. In quantum field theory, the parameter $m_B$ that
appears in the same place, the bare mass, is not observable;
it is a property of the field, defined in the absence of
interactions. A particle is conventionally identified with an
asymptotic state through the Lehmann--Symanzik--Zimmermann (LSZ)
reduction \cite{LSZ:1955}
at the pole of the full propagator.
When the field interacts, loop corrections dress the free
propagator $i/(\slashed{q}-m_B)$ into the full
propagator $i/(\slashed{q}-m_B-\Sigma_\xi(q))$, where
$\Sigma_\xi(q)$ is the self-energy in $R_\xi$ gauge; the mass
becomes inseparable from $\Sigma_\xi$, which is
itself a function of the momentum. At the pole
$\slashed{q}=m$, the pole mass $m$ is
gauge-invariant \cite{Nielsen:1975}.

The gauge invariance of the pole mass is tied to the fact that the self-energy correction is never observed in
isolation \cite{Choi:2023cqs}. In any physical process, the dressed propagator appears as an internal
line in an amplitude, where it interferes with all other
radiative corrections. The
total amplitude is gauge-invariant
and observable \cite{Abers:Lee:1973}. It has therefore been
assumed that gauge invariance is a property of the full
amplitude, not of any individual piece. The individual self-energy $\Sigma_\xi(q)$ depends on the
gauge parameter $\xi$: in $R_\xi$ gauge,
$\Sigma_P(q)\propto(1-\xi)(\slashed{q}-m)$, which vanishes
at the pole but not in general. An off-shell mass extracted from
the propagator alone is gauge-dependent and apparently
unphysical.

These facts led to the widespread conclusion that mass is a purely
on-shell concept: the pole mass $m$ is gauge-invariant, the
$\overline{\text{MS}}$ running mass $\bar{m}(\mu)$ is
gauge-invariant through the renormalization group
equation \cite{Tarrach:1981,Kronfeld:1998}, but the
momentum-dependent mass function (the quantity that actually
governs fermion propagation at arbitrary virtuality) has no
gauge-invariant definition. We show that it does: mass in quantum field theory is a well-defined gauge-invariant
function of the momentum. This is analogous to the
momentum-dependent gauge
coupling \cite{Gross:Wilczek:1973,Politzer:1973}, which has
long been observed experimentally.

Gauge-invariant off-shell Green's functions have been constructed
via the pinch
technique \cite{Cornwall:1982,Cornwall:Papavassiliou:1989,%
Binosi:Papavassiliou:2002,Binosi:Papavassiliou:2009} and the
background field
method \cite{DeWitt:1967,Denner:Dittmaier:Weiglein:1994,%
Hashimoto:Kodaira:Yasui:Sasaki:1994}. Both yield an invariant self-energy $\hat\Sigma$ that coincides with the 't~Hooft--Feynman gauge result
 $\Sigma_{\xi=1}$, establishing the gauge-invariant
building block that the present Letter extends to a
renormalized mass function.  Those constructions use
on-shell $S$-matrix elements or a modified Lagrangian;
here we work directly from the propagator.
The gauge-independent fermion self-energy was
constructed in this framework \cite{Papavassiliou:1995}; however, the
gauge-invariant mass function, renormalized and defined at
every virtuality, was not the primary object.
The Schwinger--Dyson mass function computed non-perturbatively \cite{Roberts:Williams:2021} has
remained gauge-dependent.
More broadly, the construction of a gauge-invariant charged
fermion has been an open problem since
Dirac \cite{Dirac:1955} (see \cite{Choi:2026mho}).

In this Letter we define a gauge-invariant mass function
$m(q)$ at every momentum.  The construction uses only the
Ward--Takahashi identity (WTI)~\cite{Ward:1950,Takahashi:1957}
and on-shell renormalization; it is process-independent
and works for any choice of gauge.  

\paragraph{The mass function and its properties}
The on-shell subtraction defines the renormalized mass function
\begin{equation}
  m(q) = m + \Sigma(q) - \Sigma(m)
    - (\slashed{q}-m)\frac{d\Sigma}{d\slashed{q}}(m),
  \label{eq:mren}
\end{equation}
with $m$ the pole mass and $\Sigma$ the self-energy in any fixed gauge.
The dressed propagator is
\begin{equation}
  \frac{i}{\slashed{q}-m(q)},
  \label{eq:dressed}
\end{equation}
with residue~$i$ at $\slashed{q}=m$ and $m(m)=m$.
Ultimately, the physical object is the propagator itself;
the separation into mass and kinetic term is a convention.
The mass function is well defined at every
momentum for fermions and scalars. It satisfies:
\begin{itemize}
\setlength\itemsep{0pt}
  \item[(i)] \textit{Gauge invariance}
        $\partial m(q)/\partial\xi = 0$ for all $q$, extending the Nielsen identity~\cite{Nielsen:1975} from the pole to every momentum.
  \item[(ii)] \textit{Finiteness} the on-shell subtraction
        removes all UV divergences; expressed in
        on-shell parameters, no residual
        scheme dependence remains \cite{Choi:2023cqs,Choi:2024hkd}, implying {\em scheme independence}. This is related to the {\em insensitivity to momentum-independent terms}: the subtraction structure removes all $q$-independent pieces, so no regularization-specific identity is needed.
  \item[(iii)] \textit{Process independence} $\Sigma$ is
        a two-point function depending
        only on $q$ \cite{Binosi:Papavassiliou:2002,%
        Binosi:Papavassiliou:2009,longpaper}.
  \item[(iv)] \textit{Locality}
        the coefficient of $\slashed{q}$ in the inverse
        propagator is unity for all $q$, not only at the pole,
        where it is enforced by the constant $Z_2$, but at every
        virtuality.
  \item[(v)] \textit{Field redefinition invariance}
        $m(q)$ is invariant under $\psi\to(1+h(q))\psi$.
  \item[(vi)] \textit{Decoupling} 
   heavy fields decouple from
        $\Sigma_{\text{ren}}$ as 
         $m(\slashed{q}-m)^2/M^2$ \cite{Appelquist:Carazzone:1975,Choi:2024:decoupling}.
\end{itemize}
The mass function unifies the pole mass \cite{Nielsen:1975,Tarrach:1981}, the
$\overline{\text{MS}}$ running
mass \cite{Tarrach:1981,Kronfeld:1998}, the on-shell vs.\ pole
mass prescriptions \cite{Gambino:1999,Kniehl:2002}, and the
Schwinger--Dyson mass function \cite{Roberts:Williams:2021} as
special cases or approximations of a single object.
We now derive these results.

\paragraph{Gauge symmetry requirement} We denote the free
fermion propagator by $\Sf(k) = i/(\slashed{k}-m)$ and its
inverse by $\SfI(k) = -i(\slashed{k}-m)$. The gauge-boson propagator in $R_\xi$ gauge decomposes as
\begin{equation}
  D_{\mu\nu}(\ell) = \frac{-i}{\ell^2}\!\left(g_{\mu\nu}
    - (1-\xi)\frac{\ell_\mu\ell_\nu}{\ell^2}\right)
  = D^{\xi=1}_{\mu\nu}(\ell) + D^P_{\mu\nu}(\ell),
  \label{eq:Ddecomp}
\end{equation}
where $D^{\xi=1}_{\mu\nu} = -ig_{\mu\nu}/\ell^2$ and 
$D^P_{\mu\nu} = i(1-\xi)\ell_\mu\ell_\nu/\ell^4$. 
In this work, the decomposition around $\xi=1$ is a convenience; one may equally decompose around any reference gauge $\xi_0$, as shown below. 
This induces the splitting of the one-loop fermion self-energy:
\begin{equation}
  \Sigma_\xi(q) = \Sigma_{\xi=1}(q) + \Sigma_P(q),
  \label{eq:decomp}
\end{equation}
where $\Sigma_{\xi=1}$ arises from $D^{\xi=1}$ and $\Sigma_P$ from $D^P$.
The second term $\Sigma_P$
is given by
\begin{equation}
  -i\Sigma_P(q)
  = g^2 C_2(1-\xi)\dL\frac{1}{\ell^4}
    \slashed\ell\Sf(q -  \ell)\slashed\ell,
  \label{eq:SigmaP_int}
\end{equation}
where $g$ is the gauge coupling and $C_2$ the appropriate
Casimir ($1$ in QED, $C_F$ in QCD); the color structure is
$\delta^{ab}$ and is suppressed.
Apply the tree-level WTI,
\begin{equation}
  \slashed\ell = i\SfI(q) - i\SfI(q - \ell),
  \label{eq:WTI}
\end{equation}
at each $\slashed\ell$ insertion. The double
application \cite{Binosi:Papavassiliou:2002} gives
\begin{equation}
  \slashed\ell\Sf(q-\ell)\slashed\ell
  = -\SfI(q)\Sf(q-\ell)\SfI(q)
    + 2\SfI(q) - \SfI(q-\ell).
  \label{eq:double}
\end{equation}
Upon integration, the first term is $q$-dependent and produces $(\slashed{q}-m)B_1(q^2)$.  The second is
$\SfI(q)$ times a $q$-independent integral. The third separates into $(\slashed{q}-m)$ times a
$q$-independent integral plus an odd integrand that
vanishes by Lorentz symmetry. The $q$-independent pieces do not change the Dirac
structure. The full result is
\begin{equation}
  \Sigma_P(q) = (1 - \xi)(\slashed{q}-m)B_1(q^2),
  \label{eq:SigmaP}
\end{equation}
where $B_1$ is the vector Passarino--Veltman \cite{Passarino:Veltman:1979}
two-point function with internal masses $0$ (gauge boson) and $m$ (fermion), and the
$q$-independent constants from the second and the third terms
are absorbed into $B_1$.
 
The key point is that $\Sigma_P$ is proportional to
$(\slashed{q}-m) \propto \SfI(q)$. It has the Dirac structure
of the {\em inverse free propagator,} or the kinetic term.  
This is the consequence of the WTI. It forces $\Sigma_P$
to be kinetic, which is what allows a constant $Z_2$ to
remove it. The total amplitude is independent of the gauge parameter~\cite{Abers:Lee:1973}.
The WTI guarantees that the removed piece is absorbed by the adjacent vertex, which acquires the corresponding gauge fix.
 
\paragraph{The mass function and its gauge properties}
Note that $\Sigma_P$ is proportional to $(\slashed{q}-m)$ and therefore vanishes at $\slashed{q}=m$; this is not a choice but a consequence of the Dirac structure. The part of the self-energy nonvanishing at the pole is $\Sigma_{\xi=1}$. 
The decomposition~(\ref{eq:Ddecomp}) naturally defines the mass function~(\ref{eq:mren}) at $\xi=1$, where $\Sigma_P=0$ and the vertex is $\Gamma^\mu_{\xi=1}=\gamma^\mu$.

Consider another decomposion around $\xi = \xi_0$. The self-energy $\Sigma_{\xi_0}$ splits uniquely into a part that is nonvanishing at the pole and a part that vanishes:
\begin{equation}
\Sigma_{\xi_0}(q) = \Sigma_{\xi=1}(q) + (1-\xi_0)(\slashed{q}-m)B_1(q^2).
\label{eq:split}
\end{equation}
Taking $\Sigma=\Sigma_{\xi_0}$ in~\eqref{eq:mren} gives
\begin{equation}
 m_{\xi_0}(q) = m(q)
    + (1 - \xi_0)(\slashed{q} - m)
      \bigl[B_1(q^2)-B_1(m^2)\bigr],
  \label{eq:mxi}
\end{equation}
where the field-strength renormalization $Z_2=[1-\Sigma'_{\xi_0}(m)]^{-1}$~\cite{Peskin,Sirlin:1980} has subtracted $B_1(m^2)$ but cannot remove the $q$-dependent remainder.
Since the WTI assigns the kinetic remainder to the vertex $\hat\Gamma^\mu$, the self-energy part remains the same $\hat \Sigma = \Sigma_{\xi=1}$ for {\em every $\xi_0$}; the mass function $m(q)$ is the same regardless of the computational gauge, proving property~(i).

The pinch technique~\cite{Cornwall:1982,Cornwall:Papavassiliou:1989,Papavassiliou:1990,Degrassi:Sirlin:1992,Binosi:Papavassiliou:2002,Binosi:Papavassiliou:2004,Binosi:Papavassiliou:2009,Watson:1995,Cornwall:Papavassiliou:Binosi:2011} and the background field method~\cite{DeWitt:1967,Denner:Dittmaier:Weiglein:1994,Hashimoto:Kodaira:Yasui:Sasaki:1994,Papavassiliou:1995} verified explicitly, diagram by diagram, that the WTI-mandated transfer between self-energy and vertex is consistent within physical amplitudes, arriving at $\hat\Sigma = \Sigma_{\xi=1}$. The present step is renormalization: the on-shell subtraction~\eqref{eq:mren} applied to the gauge-invariant self-energy of the pinch technique defines the mass function $m(q)$ and, as shown above, provides an elementary proof of the pole mass theorem as a corollary.

Alternatively, one may regard work with $m_{\xi_0}(q)$ at a different reference gauge $\xi=\xi_0$ and absorb the remainder into the corresponding vertex $\Gamma^\mu_{\xi_0}$.
Eq.~\eqref{eq:mxi} reveals that $m_{\xi_0}(q)$ is well-behaved (finite, with the correct pole; see below) for {\em every} $\xi_0$, not only $\xi_0=1$. The mass functions at different $\xi_0$ form a family related by kinetic terms $\propto(\slashed{q}-m)$; each member, paired with its vertex, gives the same amplitude. In this respect the present framework generalizes the pinch technique to arbitrary reference gauges; from this viewpoint, a different reference gauge $\xi_0$ corresponds to a partial pinching. The natural interpretation is provided by the Dirac dressing~\cite{Choi:2026mho}: each $\xi_0$ corresponds to a member of the dressing family, and $m_{\xi_0}(q)$ is the mass of the fermion with that dressing. This structure becomes coherent after renormalization; it is the on-shell subtraction that turns the family of bare self-energies into finite, well-defined mass functions and reveals the precise relation~\eqref{eq:mxi} among them.

All share the same pole mass $m_{\xi_0}(m)=m$.
At $\slashed{q}=m$, property~(i) reduces to the gauge invariance of the pole mass. The standard proof uses the BRST-based Nielsen identity~\cite{Nielsen:1975,Gambino:1999}; the present argument requires only the WTI, which forces all $\xi$ dependence to vanish at the pole.

\paragraph{Locality of the kinetic term} 
The one-loop corrected inverse propagator in gauge $\xi$, decomposed around reference gauge $\xi_0$ via $\Sigma_\xi = \Sigma_{\xi_0} + (\xi_0-\xi)(\slashed{q}-m)B_1(q^2)$, is
\begin{equation}
\begin{split}
 \bigl[&1-(\xi_0 - \xi)B_1(q^2)\bigr](\slashed{q}-m)
    - \Sigma_{\xi_0}(\slashed{q}) + \Sigma_{\xi_0}(m) ,\\
    &= Z(q^2)^{-1} (\slashed{q}-m) +{\cal O}\big( (\slashed{q}-m)^2 \big),
\end{split}
\end{equation}
with
\begin{equation} 
Z(q^2)^{-1}  \equiv 1-(\xi_0  - \xi)B_1(q^2)- \Sigma'_{\xi_0}(m). \label{eq:Zxi} 
\end{equation}
The kinetic coefficient is $q$-independent when $\xi=\xi_0$: for {\em any} reference gauge, computing in the corresponding gauge gives a canonical kinetic term with $Z_2 = [1-\Sigma'_{\xi_0}(m)]^{-1}$ a single number. Locality does not select a preferred $\xi_0$; it requires consistency between the reference gauge and the computational gauge.
The subtraction of $\Sigma'_{\xi_0}(m)$ in~\eqref{eq:mren} absorbs this $Z_2$ into the mass function and gives the propagator $i/(\slashed{q}-m_{\xi_0}(q))$ with no prefactor and unit residue at the pole. Conventional on-shell renormalization retains $Z_2$ as a separate wave-function factor; the present construction absorbs it into $m_{\xi_0}(q)$ at every momentum.

\paragraph{Segment locality}
The WTI relates gauge freedom of one vertex to the difference of two propagators. 
Conversely, the same identity can be read as relating two vertices to a single propagator.
That is, the gauge-parameter dependence of the propagator reduces to that of its two neighboring vertices.
For corrections where the gauge boson spans two or more external
vertices with multiple insertions in between, the detailed analysis is nontrivial and is discussed in~\cite{longpaper} (see also~\cite{Cornwall:1982,Cornwall:Papavassiliou:1989,Binosi:Papavassiliou:2002,Binosi:Papavassiliou:2009,Cornwall:Papavassiliou:Binosi:2011} for the related pinch technique constructions); however, the result is dictated by the WTI, which is itself the relation between a segment and its ending vertices. The decomposition is
{\em segment-local} \cite{longpaper}: it is determined entirely
by the gauge boson's two endpoints on the fermion line, once global cancellations take place. The exact WTI extends the segment-local
mechanism to all orders with dressed
propagators and vertices \cite{longpaper}.

It is this locality that allows the mass function to be extracted from the propagator without reference to a specific process. When a vertex is at an on-shell external end, $\SfI(q)|_{\slashed{q}=m}=0$ makes its WTI contribution trivial, confirming process independence~\cite{Cornwall:1982,Cornwall:Papavassiliou:1989,Binosi:Papavassiliou:2009}.
The mass function alone is not directly
observable; it must be paired with the gauge-invariant vertex,
defined by the same WTI, to construct a measurable amplitude. As a consequence, every internal fermion line in an arbitrary amplitude decomposes into a chain of gauge-invariant propagators $i/(\slashed{q}_i-m(q_i))$ connected by gauge-invariant vertices $\hat\Gamma^\mu$, each segment carrying a definite momentum.

In QCD, the same segment-locality applies: the non-Abelian WTI,
$\ell_\mu\Gamma^{\mu,c} = g[t^c\SfI(k) - \SfI(k')t^c]$,
triggers at the two endpoints of the internal gluon, while
external gauge bosons (photons in the Compton amplitude) are
colorless spectators. The color algebra $t^c t^c = C_F$
factors out. At higher loops, the internal gluon propagator
itself requires a gauge-invariant definition, which the
pinch technique provides through the Batalin--Vilkovisky
formalism \cite{Binosi:Papavassiliou:2002:BV}. Again, the detailed analysis is more involved, but the result is dictated by the non-Abelian WTI, which has the same segment structure; the
segment-locality of the fermion line reduces the non-Abelian
problem to the already-solved gluon sector.

\paragraph{Field redefinition invariance}
The present framework naturally accounts for the invariance~(v) under field redefinitions. A Kamefuchi--O'Raifeartaigh--Salam (KOS) transformation~\cite{KOS:1961} $\psi\to(1+h(q))\psi$ shifts $\Sigma(q)\to\Sigma(q)+h(q)(\slashed{q}-m)$. The added term is purely kinetic and therefore does not affect $m(q)$.

\paragraph{The scalar sector} The same argument applies to
scalars. Denote the free scalar propagator by
$\Delta(q)=i/(q^2-m^2)$. The scalar WTI,
\begin{equation}
  \ell\cdot(2q-\ell) =
  i\Delta^{-1}(q)-i\Delta^{-1}(q-\ell),
  \label{eq:scalarWTI}
\end{equation}
produces a difference of inverse propagators inside the loop,
exactly as the fermion WTI~\eqref{eq:WTI} does. The gauge-dependent part of the scalar self-energy,
$\Sigma_P$, arises from the same propagator
$D^P_{\mu\nu}$; the result is
$\Sigma_P(q^2)\propto(1-\xi)(q^2-m^2)$. Again,
$\Sigma_P$ is proportional to the kinetic term
$(q^2-m^2) \propto \Delta^{-1}(q)$, generating a momentum-dependent
$Z_\xi(q^2)$ that no constant scalar wave-function
renormalization $Z_\phi$ can absorb. The
resolution is the same.
The renormalized scalar mass function is
\begin{equation}
  m^2(q^2) = m^2 + \Sigma(q^2)    - \Sigma(m^2)  - (q^2-m^2)\frac{d\Sigma}{dq^2}(m^2).
  \label{eq:scalar_mren}
\end{equation}
Correction by a heavy field of mass $M$ to the scalar mass is also suppressed as $(q^2-m^2)^2/M^2$ \cite{Choi:2024:decoupling}. This is expected from the structure, because the renormalized correction should vanish at the pole $q^2=m^2$ and the mass $M$ in the propagator appears in the denominator. For the fermions, chiral symmetry suppresses the numerator and the correction is proportional to the pole mass itself $m$.

\paragraph{Physical realization}
The WTI that removes $\Sigma_P$ from the propagator simultaneously
transfers it to the vertex, yielding a gauge-invariant vertex
$\hat\Gamma^\mu$~\cite{longpaper}.
The same on-shell subtraction~\eqref{eq:mren} that defines
$m(q)$ renormalizes the vertex through $Z_1 = Z_2$
(the Ward identity); the pair $m_{\xi_0}(q)$ and $\Gamma^\mu_{\xi_0}$ gives the same amplitude for every $\xi_0$.
This has been verified explicitly in the Compton amplitude
$\gamma^*q\to\gamma^*q$ with off-shell external
fields~\cite{longpaper}. By the segment locality established above, an arbitrary amplitude reduces, up to crossing, to a chain of such Compton-type building blocks, each a gauge-invariant propagator~\eqref{eq:dressed} joined to gauge-invariant vertices $\hat\Gamma^\mu$. The all-orders exact WTI extends the
result to dressed propagators and
vertices~\cite{Slavnov:1972,Taylor:1971,Binosi:Papavassiliou:2002,Binosi:Papavassiliou:2009,longpaper}.
The extension is self-consistent order by order: at each
loop order, $D^P_{\mu\nu}$ sandwiches propagators and vertices that are already
gauge-invariant from the previous order. The WTI converts the sandwich into the dressed inverse propagator $\hat{S}_F^{-1}(q)$, so $\Sigma_P$ remains proportional to the kinetic term at every order and the
decomposition~\eqref{eq:decomp} applies recursively with
dressed objects~\cite{Choi:2025:selfsimilar}; the
Batalin--Vilkovisky framework~\cite{Binosi:Papavassiliou:2002:BV}
ensures the cancellation of higher powers of $(1-\xi)$.

\paragraph{The reality of virtual particles}
The gauge-invariant propagator~\eqref{eq:dressed} and
vertex $\hat\Gamma^\mu$ together
describe a charged particle without gauge redundancy:
$m(q)$ is as well-determined as the pole mass $m$; it is
the same function evaluated at a different momentum.

The complete one-loop Higgs mass has been computed in this
framework \cite{Choi:2023cqs,Choi:2024:decoupling,Choi:2024hkd,Choi:2025:Higgs}. The Higgs mass peaks at $0.3\%$ above the pole
value near $\sqrt{p^2} \approx 200$~GeV before decreasing quadratically in
$\sqrt{p^2}$ due to the top quark; the imaginary part reproduces the
known Higgs decay widths
exactly \cite{Choi:2025:Higgs}.
The momentum-dependent mass is now defined gauge-invariantly across
the full Standard Model: scalars, fermions, and gauge
bosons \cite{Cornwall:1982,Binosi:Papavassiliou:2009}.

The physical mass in quantum field theory is not a number but
a gauge-invariant function of the momentum.
The remaining distinction between real and virtual is
{\em not dynamical but purely kinematic}: at the pole
the propagator diverges and the particle propagates to
asymptotic distances; away from it, the same
dressed propagator~\eqref{eq:dressed} carries the same
gauge-invariant mass $m(q)$ at a different momentum.
For confined quarks the pole may not be physically
accessible, but $m(q)$ at high virtuality remains
well-defined and perturbatively computable; the
mass function is more fundamental than the pole mass.
The off-shell mass is measurable in the same sense as the pole mass: any cross section built from~\eqref{eq:dressed} and $\hat\Gamma^\mu$ isolates $m(q)-m$ at the relevant virtuality. Existing measurements of the running top-quark~\cite{CMS:running:2020} and charm-quark~\cite{HERA:charm:2018} masses already probe this dependence; the present construction supplies the gauge-invariant object that underlies them.

\begin{acknowledgments}
The author is grateful to Gian Giudice, Hyeseon Im, Taehyun Jung, Chanju Kim and Joannis Papavassiliou for discussions. This work is partly supported by the grant RS-2023-00277184 of the National Research Foundation of Korea.
The author used Claude (Anthropic) for editing and discussion during the manuscript preparation. The final content was reviewed and approved by the author, who takes full responsibility for the work.
\end{acknowledgments}


\begin{thebibliography}{99}



\bibitem{LSZ:1955}
H.~Lehmann, K.~Symanzik and W.~Zimmermann,
Nuovo Cim.\  \textbf{1} (1955), 205--225.
\doi{10.1007/BF02731765}



\bibitem{Nielsen:1975}
N.~K.~Nielsen,
Nucl.\ Phys.\ B \textbf{101} (1975), 173--188.
\doi{10.1016/0550-3213(75)90301-6}



\bibitem{Choi:2023cqs}
K.~S.~Choi,
J. Korean Phys. Soc. \textbf{84} (2024) no.8, 591-595
[erratum: J. Korean Phys. Soc. \textbf{86} (2025) no.2, 156]
doi:10.1007/s40042-024-01025-7
[arXiv:2310.00586 [hep-ph]].




\bibitem{Abers:Lee:1973}
E.~S.~Abers and B.~W.~Lee,
Phys.\ Rept.\  \textbf{9} (1973), 1--141.
\doi{10.1016/0370-1573(73)90027-6}



\bibitem{Tarrach:1981}
R.~Tarrach,
Nucl.\ Phys.\ B \textbf{183} (1981), 384--396.
\doi{10.1016/0550-3213(81)90140-8}



\bibitem{Kronfeld:1998}
A.~S.~Kronfeld,
Phys.\ Rev.\ D \textbf{58} (1998), 051501.
\doi{10.1103/PhysRevD.58.051501}
[arXiv:hep-ph/9805215 [hep-ph]].



\bibitem{Gross:Wilczek:1973}
D.~J.~Gross and F.~Wilczek,
Phys.\ Rev.\ Lett.\  \textbf{30} (1973), 1343--1346.
\doi{10.1103/PhysRevLett.30.1343}



\bibitem{Politzer:1973}
H.~D.~Politzer,
Phys.\ Rev.\ Lett.\  \textbf{30} (1973), 1346--1349.
\doi{10.1103/PhysRevLett.30.1346}



\bibitem{Cornwall:1982}
J.~M.~Cornwall,
Phys.\ Rev.\ D \textbf{26} (1982), 1453--1478.
\doi{10.1103/PhysRevD.26.1453}



\bibitem{Cornwall:Papavassiliou:1989}
J.~M.~Cornwall and J.~Papavassiliou,
Phys.\ Rev.\ D \textbf{40} (1989), 3474--3485.
\doi{10.1103/PhysRevD.40.3474}



\bibitem{Binosi:Papavassiliou:2002}
D.~Binosi and J.~Papavassiliou,
Phys.\ Rev.\ D \textbf{66} (2002), 085003.
\doi{10.1103/PhysRevD.66.085003}
[arXiv:hep-ph/0205058 [hep-ph]].



\bibitem{Binosi:Papavassiliou:2009}
D.~Binosi and J.~Papavassiliou,
Phys.\ Rept.\  \textbf{479} (2009), 1--152.
\doi{10.1016/j.physrep.2009.05.001}
[arXiv:0909.2536 [hep-ph]].



\bibitem{DeWitt:1967}
B.~S.~DeWitt,
Phys.\ Rev.\  \textbf{162} (1967), 1195--1239.
\doi{10.1103/PhysRev.162.1195}



\bibitem{Denner:Dittmaier:Weiglein:1994}
A.~Denner, G.~Weiglein and S.~Dittmaier,
Phys.\ Lett.\ B \textbf{333} (1994), 420--426.
\doi{10.1016/0370-2693(94)90162-7}
[arXiv:hep-ph/9406204 [hep-ph]].



\bibitem{Hashimoto:Kodaira:Yasui:Sasaki:1994}
S.~Hashimoto, J.~Kodaira, Y.~Yasui and K.~Sasaki,
Phys.\ Rev.\ D \textbf{50} (1994), 7066--7076.
\doi{10.1103/PhysRevD.50.7066}



\bibitem{Papavassiliou:1995}
J.~Papavassiliou,
Phys.\ Rev.\ D \textbf{51} (1995), 856--861.
\doi{10.1103/PhysRevD.51.856}
[arXiv:hep-ph/9410385 [hep-ph]].



\bibitem{Roberts:Williams:2021}
C.~D.~Roberts and S.~M.~Schmidt,
Eur.\ Phys.\ J.\ ST \textbf{229} (2020), 3319--3340.
\doi{10.1140/epjst/e2020-000064-6}
[arXiv:2006.08782 [hep-ph]].



\bibitem{Dirac:1955}
P.~A.~M.~Dirac,
Can.\ J.\ Phys.\  \textbf{33} (1955), 650--660.
\doi{10.1139/p55-081}


\bibitem{Ward:1950}
J.~C.~Ward,
Phys.\ Rev.\  \textbf{78} (1950), 182.
\doi{10.1103/PhysRev.78.182}



\bibitem{Takahashi:1957}
Y.~Takahashi,
Nuovo Cim.\  \textbf{6} (1957), 371--375.
\doi{10.1007/BF02826513}

\bibitem{Choi:2024hkd}
K.~S.~Choi,
[arXiv:2410.21118 [hep-ph]].

\bibitem{longpaper}
K.-S.~Choi and H.~Im,
``Gauge-invariant renormalized off-shell mass'' (companion paper), to appear.




\bibitem{Appelquist:Carazzone:1975}
T.~Appelquist and J.~Carazzone,
Phys.\ Rev.\ D \textbf{11} (1975), 2856--2861.
\doi{10.1103/PhysRevD.11.2856}



\bibitem{Choi:2024:decoupling}
K.-S.~Choi,
[arXiv:2408.06406 [hep-ph]].



\bibitem{Gambino:1999}
P.~Gambino and P.~A.~Grassi,
Phys.\ Rev.\ D \textbf{62} (2000), 076002.
\doi{10.1103/PhysRevD.62.076002}
[arXiv:hep-ph/9907254 [hep-ph]].



\bibitem{Kniehl:2002}
B.~A.~Kniehl, F.~Madricardo and M.~Steinhauser,
Phys.\ Rev.\ D \textbf{62} (2000), 073010.
\doi{10.1103/PhysRevD.62.073010}
[arXiv:hep-ph/0005060 [hep-ph]].



\bibitem{Passarino:Veltman:1979}
G.~Passarino and M.~J.~G.~Veltman,
Nucl.\ Phys.\ B \textbf{160} (1979), 151--207.
\doi{10.1016/0550-3213(79)90234-7}



\bibitem{Peskin}
M.~E.~Peskin and D.~V.~Schroeder,
Addison-Wesley, Reading, 1995.



\bibitem{Papavassiliou:1990}
J.~Papavassiliou,
Phys.\ Rev.\ D \textbf{41} (1990), 3179--3191.
\doi{10.1103/PhysRevD.41.3179}



\bibitem{Degrassi:Sirlin:1992}
G.~Degrassi and A.~Sirlin,
Phys.\ Rev.\ D \textbf{46} (1992), 3104--3116.
\doi{10.1103/PhysRevD.46.3104}



\bibitem{Binosi:Papavassiliou:2004}
D.~Binosi and J.~Papavassiliou,
J.\ Phys.\ G \textbf{30} (2004), 203--234.
\doi{10.1088/0954-3899/30/2/017}
[arXiv:hep-ph/0301096 [hep-ph]].



\bibitem{Watson:1995}
N.~J.~Watson,
Phys.\ Lett.\ B \textbf{349} (1995), 155--164.
\doi{10.1016/0370-2693(95)00228-G}



\bibitem{Cornwall:Papavassiliou:Binosi:2011}
J.~M.~Cornwall, J.~Papavassiliou and D.~Binosi,
Cambridge University Press, Cambridge, 2011.



\bibitem{Sirlin:1980}
A.~Sirlin,
Phys.\ Rev.\ D \textbf{22} (1980), 971--981.
\doi{10.1103/PhysRevD.22.971}



\bibitem{Binosi:Papavassiliou:2002:BV}
D.~Binosi and J.~Papavassiliou,
Phys.\ Rev.\ D \textbf{66} (2002), 025024.
\doi{10.1103/PhysRevD.66.025024}
[arXiv:hep-ph/0204128 [hep-ph]].



\bibitem{Slavnov:1972}
A.~A.~Slavnov,
Theor.\ Math.\ Phys.\  \textbf{10} (1972), 99--107.
\doi{10.1007/BF01090719}



\bibitem{Taylor:1971}
J.~C.~Taylor,
Nucl.\ Phys.\ B \textbf{33} (1971), 436--444.
\doi{10.1016/0550-3213(71)90297-5}



\bibitem{Choi:2025:selfsimilar}
K.-S.~Choi,
[arXiv:2502.19300 [hep-th]].







\bibitem{Choi:2025:Higgs}
K.-S.~Choi,
[arXiv:2506.18667 [hep-ph]].



\bibitem{CMS:running:2020}
CMS Collaboration,
Phys.\ Lett.\ B \textbf{803} (2020), 135263.
\doi{10.1016/j.physletb.2020.135263}
[arXiv:1909.09193 [hep-ex]].



\bibitem{HERA:charm:2018}
A.~Gizhko \textit{et~al.} [H1 and ZEUS],
Phys.\ Lett.\ B \textbf{775} (2017), 233--243.
\doi{10.1016/j.physletb.2017.09.055}
[arXiv:1705.08863 [hep-ex]].

\bibitem{Choi:2026mho}
K.~S.~Choi,
[arXiv:2603.13684 [hep-th]].



\bibitem{KOS:1961}
S.~Kamefuchi, L.~O'Raifeartaigh and A.~Salam,
Nucl.\ Phys.\  \textbf{28} (1961), 529--549.
\doi{10.1016/0029-5582(61)90056-6}



\end{thebibliography}
\end{document}